# Performance of Quantum Data Transmission Systems in the Presence of Thermal Noise

G. Cariolaro, *Life Member, IEEE* and G. Pierobon *Life Member, IEEE*

*Abstract*— In the literature the performance of quantum data transmission systems is usually evaluated in the absence of thermal noise. A more realistic approach taking into account the thermal noise is intrinsically more difficult because it requires dealing with Glauber coherent states in an infinite–dimensional space. In particular, the exact evaluation of the optimal measurement operators is a very difficult task, and numerical approximation is unavoidable. The paper faces the problem by approximating the $P$–representation of the noisy quantum states with a large number of terms and applying to them the square root measurement (SRM) approach. Comparisons with cases where the exact solution are known show that the SRM approach gives quite accurate results. As application, the performance of quadrature amplitude modulation (QAM) and phase shift keying (PSK) systems is considered. In spite of the fact that the SRM approach is not optimal and overestimates the error probability, also in these cases the quantum detection maintains its superiority with respect to the classical homodyne detection.

## I. INTRODUCTION

Transmission of information through a quantum channel is mainly affected by an uncertainty which is intrinsically related to the quantum mechanics laws. In the language of classical optical systems, this uncertainty corresponds to the so called *shot noise*. Another reason of uncertainty is the presence of *thermal noise*, as in classical systems. Beginning from 1970's, a lot of research work has been devoted to the *quantum detection problem* [10], that may be summarized in the following terms. The transmitter sends a quantum signal through a quantum channel, which forces the receiver to assume one among a finite number of states. The detector tries to guess the state by an adequate set of quantum measurements and the problem arises of finding the measurement set which optimizes the detection, according to some predefined fidelity criterion (usually the minimum error probability). Necessary and sufficient conditions for the optimal measurement set have been found in pioneering papers by Holevo [11] and by Yuen *et al.* [15].

Unfortunately, even though the optimal measurement set is completely characterized, analytical closed–form solutions for the measurement set, indeed for the minimum error probability, are not available, in general. Then, it is needed to resort to a numerical evaluation based on convex semidefinite programming [6]. However, under some rotation symmetry constraint, a simple measurement, introduced by Hausladen *et al.* [8] and known as *square root measurement* (SRM), turns out to be optimum. The SRM has the remarkable advantage that it is straightforwardly evaluated starting from the possible states. Moreover, also when it is not optimal, the SRM often gives "pretty good" upper bounds on the performance of optimal detectors.

Our paper starts just from these important results on SRM for studying quantum data transmission systems in the presence of thermal noise. After the pioneering work in [10], the problem of quantum detection in a noisy environment has received scarce attention in the literature. To the best of our knowledge, only a correspondence by Sasaki *et al.* [13] on quantum on–off keying (OOK) and a technical report [14] on quantum PSK attempt to afford an approximate analysis of quantum detection of coherent states. This delay is due to the difficulties of performing efficient approximations in the numerical performance evaluation. On the other hand, exploiting the new perspectives open by the extension of the SRM to mixed states [6], we apply this approach to a quantum noisy channel according to the Glauber theory on coherent states.

The paper is organized as follows. In Section II we review quantum detection fundamentals. In Sections III the SRM techniques are recalled and in Section IV the key problem of the finite-dimensional factorization of the Glauber representation of noisy states is discussed. Finally, in Sections V and VI the SRM approach is applied to QAM and PSK quantum systems (the same systems considered by Kato *et al.* [12] in the absence of thermal noise).

## II. QUANTUM DATA TRANSMISSION SYSTEM

In this section we recall some basic facts about quantum detection following the scheme of Fig. 1. For a detailed treatment the reader is referred to [10] and for a more recent survey to [2].

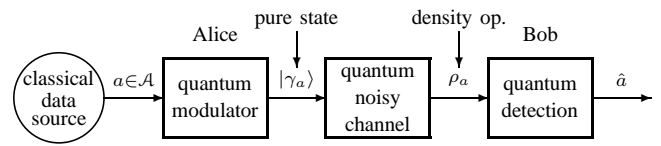

Fig. 1 — Model of a quantum data transmission system.

### A. General Model

A classical source emits a symbol $a$ drawn from a finite alphabet $\mathcal{A} = \{0, 1, \ldots, m-1\}$ with a given *prior probability distribution* $q_i = \mathrm{P}\,[a = i]$ , $i = 0, \ldots, m-1$. On the basis of the symbol $a$ emitted by the source, the transmitter sends a quantum state $|\gamma_a\rangle$ through a quantum channel (e.g., a laser pulse through an optical fiber). As a consequence, the received state is one of $m$ possible quantum states and the detection device performs a set of measurements in order to guess the received state and consequently the original symbol.

In the ideal case, i.e. neglecting thermal noise, a set of $m$ pure states is seen by the receiver, which are a replica of the transmitted states $|\gamma_a\rangle$. In the presence of thermal noise the received states become noisy (or mixed) and are described by a set of *density operators* $\rho_i$, $i = 0, \ldots, m-1$, which are Hermitian, positive semidefinite (PSD) and have unit trace, $\mathrm{Tr}(\rho_i) = 1$. The description through density operators represents the general case, since it is comprehensive of the pure state case, in which $\rho_i$ reduces to the rank–one operator $\rho_i = |\gamma_i\rangle\langle\gamma_i|$.

### B. Quantum Detection Theory. Available Results.

Quantum theory postulates that a detection device for choosing among the possible states is given by a *positive operator valued measurement* (POVM), i.e. a set of $m$ operators $\Pi_0, \ldots, \Pi_{m-1}$ that are Hermitian, PSD and resolve the identity operator of the Hilbert space $\mathcal{H}$, namely $\sum_{i=0}^{m-1} \Pi_i = I_{\mathcal{H}}$. Then, the probability that the detection system reveals the state $j$, provided that the state density operator is $\rho_i$, is given by

$$p(j|i) = \mathrm{Tr}(\rho_i \Pi_j), \qquad i, j = 0, \ldots, m-1. \quad (1)$$

In particular, the probability of correct detection becomes

$$P_c = \sum_{i=0}^{m-1} q_i \, p(i|i) = \sum_{i=0}^{m-1} q_i \, \mathrm{Tr}(\rho_i \Pi_i). \quad (2)$$

For pure states, that is with $\rho_i = |\gamma_i\rangle\langle\gamma_i|$, rank–one POVMs of the form $\Pi_j = |\mu_j\rangle\langle\mu_j|$ can be used, where $|\mu_j\rangle$ are called *measurement vectors*. Then, (1) reduces to $p(j|i) = |\langle\gamma_i|\mu_j\rangle|^2$.

The optimization of the detection scheme reduces to finding the POVM operators $\Pi_i$ that maximize $J = \sum_{i=0}^{m-1} \mathrm{Tr}(q_i \rho_i \Pi_i)$ under the constraints that the $\Pi_i$ are PSD and resolve the identity $I_{\mathcal{H}}$. The maximum of $J$ is the optimal probability of correct detection. Clearly, this is a convex semidefinite programming problem in the real space of the Hermitian operators.

On the other hand, analytical closed–form results are available only for the particular class of pure states exhibiting the so called *geometrically uniform symmetry* (GUS). In this case Ban *et al.* [1] have shown that the optimal POVMs are given by the SRM. This particular solution has been thoroughly discussed by Eldar and Forney [4]. Recently the SRM approach has been extended to mixed states by Eldar *et al.* [6].

## III. The SRM Techniques

In this section we review the SRM techniques, having in mind that our final application will be the optical quantum transmissions, where quantum states and density operators should be formulated according to the Glauber theory (see Section IV). The SRM technique is here considered in the general case of mixed states, following Eldar, Megretski and Verghese [6].

### A. General Formulation

We start from a constellation of $m$ density operators $\rho_0, \ldots, \rho_{m-1}$ in an $n$-dimensional Hilbert space $\mathcal{H}$. The key of the SRM approach is the factorization of each density operator in the form $\rho_i = \gamma_i \gamma_i^*$ for some complex matrices $\gamma_i$, e.g., via the eigendecomposition of $\rho_i$. The factorization is not unique, but the ambiguity is irrelevant for the quantum decision. If $\rho_i$ has rank $r_i \le n$, the factor $\gamma_i$ can be chosen to have dimensions $n \times r_i$. In [6] $\gamma_i$ is referred as a factor of $\rho_i$, but, more specifically, we call $\gamma_i$ a *state factor*. Since the $i$-th optimal measurement operator can be chosen with rank not greater than the rank of $\rho_i$ [5], the search can be confined to POVMs of the form $\Pi_i = \mu_i \mu_i^*$ where $\mu_i$ are $n \times r_i$ complex matrices. We refer to $\mu_i$ as *measurement factors*.

The (generalized) *state* and the *measurement matrices* are obtained by storing the corresponding $n \times r_i$ factors as block–column vectors, namely $\Gamma = [\gamma_0, \gamma_1, \ldots, \gamma_{m-1}]$ and $M = [\mu_0, \mu_1, \ldots, \mu_{m-1}]$. The dimensions of both $\Gamma$ and $M$ are $n \times k$ with $k = r_0 + \cdots + r_{m-1}$.

From the state matrix $\Gamma$ we form the $k \times k$ Gram matrix $G = \Gamma^* \Gamma$ and also the $n \times n$ matrix $T = \Gamma \Gamma^*$, sometimes called Gram matrix.

At this point the SRM method is used to provide the measurement matrix $M$. The first step is the eigendecompositions of $G$ and $T$, namely $G = V \Lambda_G V^*$, $T = U \Lambda_T U^*$, where $U$, $V$ are unitary and $\Lambda_G$, $\Lambda_T$ are diagonal. Note that $G$ and $T$ are both PSD with the same rank $r$ and have the same set of positive eigenvalues [4]. From the eigendecompositions we can find the inverse of the square root of $G$ and $T$ as $G^{-1/2} = V \Lambda_G^{-1/2} V^*$ and $T^{-1/2} = U \Lambda_T^{-1/2} U^*$, where the inverses must be intended in the generalized Moore–Penrose sense [4]. Finally, the measurement matrix is given by $M = T^{-1/2} \Gamma$.

An alternative evaluation of $M$ is obtained through the Gram matrix. In fact, it can be shown [4] using the singular–value decomposition, that the measurement matrix is also given by $M = \Gamma \, G^{-1/2}$. Then, the evaluation of the transition probabilities follows from (1), namely

$$\begin{aligned}
p(j|i) &= \mathrm{Tr}(\rho_i \Pi_j) = \mathrm{Tr}(\gamma_i \gamma_i^* \mu_j \mu_j^*) \\
&= \mathrm{Tr}(\mu_j^* \gamma_i \gamma_i^* \mu_j) = \mathrm{Tr}(B_{ji} B_{ji}^*)
\end{aligned} \quad (3)$$

where $B_{ji}$ is the $(j, i)$–th block of the matrix $M^* \Gamma = G^{-1/2} \Gamma^* \Gamma = G^{1/2}$. Then, for the evaluation of $p(j|i)$ we have to partition $G^{1/2}$ into blocks. Finally, the probability of correct decision becomes

$$P_c = \frac{1}{m} \sum_{i=0}^{m-1} \mathrm{Tr}(B_{ii}^* B_{ii}).$$

This twofold possibility, via $T^{-1/2}$ or $G^{\pm 1/2}$, is very important for an efficient computation, particularly when $k < n$.

### B. SRM with Geometrically Uniform Symmetry

The SRM is simplified and provides peculiar properties if the state constellation exhibits the *geometrically uniform symmetry* (GUS), that is if there exists a unitary operator[1] $S$, such that $S^m = I_{\mathcal{H}}$ and $\rho_i = S^i \rho_0 S^{-i}$. The operator $S$ and the density operator $\rho_0$ are said the *generating operator* and the *generating density* of the constellation, respectively. For the mixed states the factorization $\rho_0 = \gamma_0 \gamma_0^*$ leads to the form $\gamma_i = S^i \gamma_0$. Note that with GUS all the state factors $\gamma_i$ have the same rank $h$ as $\gamma_0$.

Since $S$ is unitary, its eigendecomposition has the form $S = Y \Lambda Y^* = \sum_{j=0}^{n-1} \lambda_j |y_j\rangle\langle y_j|$, where $Y$ is a unitary matrix of order $n$ and the eigenvalues $\lambda_j$ collected in the diagonal matrix $\Lambda$ have unit amplitude. Moreover, because of $S^m = I_{\mathcal{H}}$, the diagonal matrix has the form $\Lambda = \text{diag}\left[W_m^{r_0}, \ldots, W_m^{r_{n-1}}\right]$ where $W_m = e^{i2\pi/m}$ and the exponents $r_i$ are suitable integers with $0 \le r_i < m$. By collecting the terms with equal eigenvalues in the eigendecomposition, one gets $S = \sum_{k=0}^{m-1} W_m^k Y_k$, where $Y_k$ are projector operators, so that $Y_h Y_k = Y_h \delta_{hk}$.

The Gram matrix $G = \Gamma^* \Gamma$, of order $mh$, is formed by the blocks of order $h$

$$G_{rs} = \gamma_r^* \gamma_s = \gamma_0^* S^{s-r} \gamma_0$$
$$= \sum_{k=0}^{m-1} W_m^{k(s-r)} \gamma_0^* Y_k \gamma_0 = \frac{1}{m} \sum_{k=0}^{m-1} W_m^{k(s-r)} D_k \ ,$$

where $D_k = m \gamma_0^* Y_k \gamma_0$. Since $G_{rs}$ depends only on the difference $r - s$ mod $m$ the Gram matrix is *block circulant*. But, the important point is that this property yields an explicit decomposition for $G$, namely

$$G = V_{m,h} D V_{m,h}^* \qquad (4)$$

where $D = \text{diag}\{D_0, \ldots, D_{m-1}\}$ and $V_{m,h}$ is the $hm \times hm$ matrix $V_{m,h} = \|(1/\sqrt{m}) W_m^{rs} I_h\|$ with $I_h$ identity matrix of order $h$. As a consequence, the diagonal blocks of the matrix are given by the Discrete Fourier Transform (DFT) of the first "block row" of the Gram matrix $G$, namely $D_i = \sum_{s=0}^{m-1} G_{0s} W_m^{-is}$.

Note that (4) is not a standard eigendecomposition because the blocks $D_i$ are not diagonal matrices. To find the square root of $G$ we have to evaluate the square root of $D$ with an eigendecomposition of each block $D_i$. Since these are PSD Hermitian square matrices, their square roots matrices $D_i^{\pm 1/2}$ can be calculated to construct $D^{\pm 1/2} = \text{diag}\left[D_0^{\pm 1/2}, \ldots, D_{m-1}^{\pm 1/2}\right]$. Finally, we obtain $G^{\pm 1/2} = V_{m,h} D^{\pm 1/2} V_{m,h}^*$ whose $(r, s)$ block is given by

$$[\,G^{\pm 1/2}\,]_{rs} = \frac{1}{\sqrt{m}} \sum_{i=0}^{m-1} W_m^{(s-r)i} D_i^{\pm 1/2} \ . \qquad (5)$$

Now, the probabilities $p(j|i)$ are obtained by applying (3) with $B_{ji} = (G^{1/2})_{ji}$. In particular, $p(i|i)$ are independent of

$i$ and give the probability of correct detection. The explicit result is

$$P_c = p(i|i) = \text{Tr}\left[B_{ii}^2\right] = \text{Tr}\left[(G^{1/2})_{ii}\right]$$
$$= \frac{1}{m} \text{Tr}\left[\left\{\sum_{k=0}^{m-1} D_k^{1/2}\right\}^2\right] . \qquad (6)$$

Finally, we recall that the SRM is optimal for GUS pure states, but not for GUS mixed states, at least in general. In particular, the sufficient condition for optimality given in [6] fails in all our examples of application.

## IV. Signal and Noise in Quantum Optical Communications

In this section we recall the quantum environment for the signal and thermal noise in optical communications. The correct settlement is provided by the celebrated Glauber theory [7], which represents signal and noise in an infinite dimensional Hilbert space.

### A. Representation of Coherent States

The quantum model of a *coherent state* representing a monochromatic electromagnetic radiation produced by a laser is formulated in an infinite dimensional Hilbert space $\mathcal{H}$ equipped with an orthonormal basis $|n\rangle, n = 0, 1, 2, \ldots$, where $|n\rangle$ are called *number eigenstates*. Each state $|n\rangle$ is said to contain exactly $n$ photons. In this context the Glauber representation of a single radiation mode is given by the ket

$$|\gamma\rangle = e^{-\frac{1}{2}|\gamma|^2} \sum_{n=0}^{\infty} \frac{\gamma^n}{\sqrt{n!}} |n\rangle \qquad (7)$$

where $\gamma$ is the complex envelope that specifies the mode. Thus, for each $\gamma \in \mathbb{C}$ a coherent state (or Glauber state) is defined; in particular, the state $|0\rangle$ obtained with $\gamma = 0$ represents the *ground state*. The probability of obtaining exactly $m$ photons is governed by the Poisson distribution $p(m|\gamma) = |\langle m|\gamma\rangle|^2 = \exp(-|\gamma|^2) \ |\gamma|^{2m}/m!$ with mean $|\gamma|^2$. Hence, $N_\gamma = |\gamma|^2$ represents the *average number of photons* when the system is in the coherent state $|\gamma\rangle$. We recall that the Glauber states are not orthogonal, since the inner product of two Glauber states is given by

$$\langle \alpha | \beta \rangle = e^{-\frac{1}{2}(|\alpha|^2 + |\beta|^2 - 2\alpha^*\beta)} \ . \qquad (8)$$

The representation (7) is only valid when the receiver observes a pure state with a known parameter $\gamma$, which in the context of communications may be regarded as the *signal*. In the presence of thermal (or background) noise the signal $\gamma$ becomes uncertain and must be represented through a density operator. The Glauber theory [7][10] states that the density operator is given by

$$\rho(\gamma) = \frac{1}{\pi N} \int_{\mathbb{C}} \exp\left(-\frac{|\alpha - \gamma|^2}{N}\right) |\alpha\rangle\langle\alpha| \, d\alpha \qquad (9)$$

that is by a continuous mixture of coherent states. In (9) the parameter $N$ represents the *average number of photons* associated with the thermal noise; it is given by $N =$ $1/[\exp{(h\nu/kT_0)} - 1]$, with $h$ Planck's constant, $k$ Boltzmann's constant, $\nu$ optical frequency and $T_0$ absolute temperature of the receiver. Hence, the representation of "signal plus noise" depends only on the two parameters: 1) $\gamma \in \mathbb{C}$, which determines the nominal coherent state $|\gamma\rangle$, and 2) $\mathcal{N}$ representing the average number of noise photons. When $\mathcal{N} = 0$, that is, in the absence of noise, relation (9) degenerates into the pure state density operator $\rho(\gamma) = |\gamma\rangle\langle\gamma|$.

---

[1]The GUS can be generalized over a group of unitary operators [6], but in our applications this generalization is not needed.

### B. Discretization of the Density Operators

An infinite matrix representation $||\rho_{mn}||$ of the density operator (9) is obtained in terms of the orthonormal basis of the number eigenstates $|n\rangle$, namely $\rho_{mn} = \langle m|\rho(\gamma)|n\rangle$, and the expression of the $m\,n$ entry is [9]

$$\rho_{mn}(\gamma) = (1-v)v^n \sqrt{\frac{m!}{n!}} \left(\frac{\gamma^*}{\mathcal{N}}\right)^{n-m} \cdot$$
$$\cdot e^{-(1-v)|\gamma|^2} L_m^{n-m}\left(-\frac{|\gamma|^2}{\mathcal{N}(\mathcal{N}+1)}\right) \quad (10)$$

where $0 \leq m \leq n$, $\gamma \neq 0$, $v = \mathcal{N}/(1+\mathcal{N})$ and $L_m^{n-m}(x)$ are the generalized Laguerre polynomials. The entries for $m > n$ are obtained by the symmetry $\rho_{nm}(\gamma) = \rho_{mn}^*(\gamma)$. The matrix is infinite dimensional and not diagonal. The diagonal elements $\rho_{mm}(\gamma)$ give the probabilities of obtaining exactly $m$ photons when the quantum system is in the noisy state $\rho(\gamma)$ [9]. From (10) we have

$$p_L(m) = \rho_{mm}(\gamma)$$
$$= (1-v)v^m\, e^{-(1-v)N_\gamma}\, L_m\left((1-v)^2 N_\gamma/v\right) \quad (11)$$

which represents the *Laguerre distribution* ($L_m(x) = L_m^0(x)$ are the ordinary Laguerre polynomials). The mean and the variance of distribution (11) are $N_\gamma + \mathcal{N}$ and $N_\gamma + 2N_\gamma\mathcal{N} + \mathcal{N}(\mathcal{N}+1)$, respectively. For the ground state $|\gamma\rangle = |0\rangle$ the above expressions degenerate. The matrix representation becomes diagonal, namely

$$\rho_{mn}(0) = \delta_{mn}(1-v)v^n \quad (12)$$

and the distribution becomes *geometrical*: $p_G(m) = \rho_{mm}(0) = (1-v)v^m$. The infinite dimension matrix $||\rho_{mn}(\gamma)||, 0 \leq m, n < \infty$ gives a correct representation of the density operator. But, for the SRM, which is based on

eigendecompositions, we need a finite dimensional approximation by a truncation to $N_\epsilon$ terms. For the choice of $N_\epsilon$ to get a given accuracy we follow the *quasi–unitary trace criterion*, which is based on the fact that a density operator has a unitary trace. Then, we choose $N_\epsilon$ as the smallest integer such that

$$\sum_{m=0}^{N_\epsilon-1} \rho_{mm}(\gamma) = \sum_{m=0}^{N_\epsilon-1} p_L(m) \geq 1 - \epsilon$$

where $\epsilon$ is the required accuracy. Thus, for a given $\epsilon$, $N_\epsilon$ can be evaluated using the Laguerre distribution (11).

### C. Factorization of the Density Operators

Once established the finite $N_\epsilon \times N_\epsilon$ approximation of the density operators, for the SRM we need a factorization of the form $\rho(\alpha) = \gamma(\alpha)\gamma^*(\alpha)$ for a convenient matrix $\gamma(\alpha)$, which we call state factor. This is obtained from the eigendecomposition of $\rho(\alpha)$, namely

$$\rho(\alpha) = \sum_{i=1}^r \lambda_i^2\, |u_i\rangle\langle u_i| = U_r \Lambda_r^2 U_r^*$$

where $r$ is the rank of $\rho(\alpha)$, $U_r$ is $N_\epsilon \times r$ and collects the eigenvectors $|u_i\rangle$ corresponding to the $r$ positive eigenvalues $\lambda_i^2$, which are assumed in a decreasing order, and $\Lambda_r^2$ is $r \times r$ diagonal collecting the $\lambda_i^2$. Hence, $\gamma(\alpha) = U_r \Lambda_r$ is a correct factor of $\rho(\alpha)$. (For $\alpha = 0$ (ground state) the factorization is immediate since $\rho(0)$ is diagonal and from (12) we find $\gamma(0) = \sqrt{\rho(0)} = ||\delta_{mn}\sqrt{(1-v)v^n}||$.)

A critical point in the numerical evaluation is the choice of the rank $r$, given by the number of the numerically relevant positive eigenvalues. To clarify the problem we develop a specific case: $\alpha = \sqrt{5}$, $N_\alpha = 5$, $\mathcal{N} = 0.1$, $\epsilon = 10^{-5} \rightarrow N_\epsilon = 20$. Now, in theory $\rho(\alpha)$ has a full rank $r = N_\epsilon$, as we can see from the list of its eigenvalues obtained with a great accuracy

$0.150285$, $0.00231095$, $0.0000353779$, $5.2072510^{-7}$, $7.2487410^{-9}$
$9.4515710^{-11}$, $1.1460310^{-12}$, $\ldots$, $1.344510^{-19}$, $-4.1356410^{-20}$
$-3.1086710^{-21}$, $2.3218610^{-21}$, $-1.3763110^{-22}$, $3.0177510^{-25}$

but in practice we can limit to take only the first 3 eigenvalues, neglecting the remaining, which means that we assume as a "practical" rank $r = 3$. As a check, the reconstruction of $\rho(\alpha)$ obtained in such a way assures an accuracy $< 10^{-7}$. To find the "practical" rank in general we consider the reconstruction error $\Delta\rho = \rho - \gamma_r \gamma_r^*$, where the factor $\gamma_r$ is obtained by considering only $r$ eigenvalues. Then, we can evaluate the maximum error or the mean square error (m.s.e.) as a function of $r$ and we choose $r = N_\nu$ to achieve a given accuracy $\nu$.

Table 1: Values of $N_\epsilon$ and $N_\nu$ of the dimensions of state factors for some values of the average number of photons $N_\gamma$ and of the thermal noise parameter $\mathcal{N}$.

| $N_\gamma \rightarrow$ | 0.5 | | 1.0 | | 5 | | 10 | | 15 | | 25 | |
|---|---|---|---|---|---|---|---|---|---|---|---|---|
| | $N_\epsilon$ | $N_\nu$ | $N_\epsilon$ | $N_\nu$ | $N_\epsilon$ | $N_\nu$ | $N_\epsilon$ | $N_\nu$ | $N_\epsilon$ | $N_\nu$ | $N_\epsilon$ | $N_\nu$ |
| $\mathcal{N} = 0.001$ | 7 | 2 | 10 | 2 | 21 | 2 | 31 | 2 | 40 | 2 | 57 | 2 |
| $\mathcal{N} = 0.01$ | 7 | 3 | 9 | 3 | 20 | 2 | 30 | 2 | 39 | 2 | 55 | 2 |
| $\mathcal{N} = 0.1$ | 9 | 4 | 11 | 4 | 22 | 4 | 32 | 4 | 41 | 4 | 57 | 4 |
| $\mathcal{N} = 1.0$ | 21 | 12 | 24 | 12 | 38 | 11 | 51 | 11 | 62 | 10 | 81 | 10 |
| $\mathcal{N} = 2.0$ | 33 | 18 | 36 | 18 | 52 | 17 | 66 | 17 | 78 | 16 | 99 | 16 |
| $\mathcal{N} = 3.0$ | 45 | 24 | 49 | 24 | 67 | 23 | 83 | 22 | 97 | 21 | 121 | 20 |

In Table 1 we give a collection of $N_\epsilon$ and $N_\nu$ for some values of $N_\gamma$ and $N$, obtained with the accuracies $\epsilon = \nu = 10^{-5}$. $N_\nu$ was obtained considering the m.s.e.

## V. APPLICATION TO QAM MODULATION

The $m$–QAM constellation is defined starting from the auxiliary alphabet $\mathcal{A}_L = \{-(L-1)+2(i-1)|\ i=1,2,\ldots,L\}$ with $L = 2,3,\ldots$ and is given by the $m$ Glauber states

$$|\gamma_{uv}\rangle = |\Delta(u+iv)\rangle, \qquad u,v \in \mathcal{A}_L \tag{13}$$

with $m = L^2$. This constellation has not the GUS and therefore the SRM must be applied in the general form. In (13) $\Delta$ is a scale factor, which determines the average number of signal photons, specifically

$$N_s = \frac{2}{3}(L^2-1)\Delta^2 = \frac{2}{3}(m-1)\Delta^2 \ . \tag{14}$$

For instance, for the 16–QAM we find $N_s = 10\Delta^2$.

In the case of pure states the first step is the evaluation of the Gram matrix $G$, whose elements are given by the inner products[2]

$$\begin{aligned} \langle\gamma_{uv}|\gamma_{u'v'}\rangle &= \langle\Delta(u+iv)|\Delta(u'+iv')\rangle \\ &= \exp\{-\tfrac{1}{2}\Delta^2[(u'-u)^2 + (v'-v)^2 - 2i(u'v-v'u)]\} \\ &\qquad u,v,u',v' \in \mathcal{A}_L \ . \end{aligned}$$

Then, the eigendecomposition $G = V\Lambda_G V^*$, the evaluation of $G^{1/2}$ and of the probabilities can be carried out without approximation and with a low computational complexity since the dimensions involved are only $m \times m$ [12].

### A. Application of SRM in the Presence of Noise

In the presence of thermal noise the only problem is the management of approximations since the density operators must be approximated by finite–dimensional matrices, as discussed in Section IV. In the QAM format the average number of photons $N_\gamma = |(u+iv)\Delta|^2$, is not uniform over the constellation, varying from $N_\gamma = 2\Delta^2$ for the inner symbols to $N_\gamma = 2(L-1)^2\Delta^2$ for the corner symbols. The reduced dimensions of the Hilbert space $n = N_\epsilon$ must be chosen considering the maximum $N_{\gamma,\max} = 2(L-1)^2\Delta^2$. Then, assuming $N_s$ as fundamental parameter, for the choice of $N_\epsilon$ we have to consider that

$$N_{\gamma\max} = 3[(L-1)^2/(L^2-1)]\,N_s = 3[(L-1)/(L+1)]\,N_s \ .$$

For instance, for 16–QAM we find $N_{\gamma\max} = 1.8N_s$.

We sketch an example to illustrate the dimensions involved in the 16–QAM. With $N_s = 4$ and $N = 0.1$ we find $N_{\gamma\max} = 7.2$ and we choose $N_\epsilon = 40$ assuring an accuracy $\epsilon = 10^{-7}$. The dimensions of the $\rho_i$ are $40 \times 40$ and they are factored into matrices $\gamma_i$ of dimensions $40 \times 8$. The dimensions of $\Gamma$, $T$ and $G$ are $40 \times 128$, $40 \times 40$, $128 \times 128$, respectively. So, it is more efficient to compute $T^{-1/2}$ rather than $G^{\pm 1/2}$. With the above choices we find the following diagonal transition probabilities: 0.875749 for inner states, 0.916501 for side states, 0.947767 for corner states, and the average error probability is $P_e = 0.08587$.

The SRM approach has been applied systematically to evaluate the error probability $P_e$ in the 16–QAM systems following the steps outlined above. The results are illustrated in Fig. 2, where $P_e$ is plotted versus the average number of photons per symbol $N_s$ for some values of the thermal noise parameter $N$. In particular, the curve for $N = 0$, which refers to the absence of thermal noise, was checked with the results obtained with pure states and a perfect agreement has been found. To assure an overall accuracy of $\epsilon = 10^{-7}$, the dimensions of the Hilbert space have been set to the value $n = N_\epsilon = 130$.

In Fig. 2 the performance of the SRM quantum detector is also compared with the performance of the classical homodyne detector, for which a Gaussian additive model with SRN=$4N_s/(1+2N)$ results.[3] In the absence of thermal noise an improvement of about 3 dB over homodyne detection is confirmed. As it was expected, this improvement rapidly reduces as thermal noise increases. (For a comparison with optimum detection see the end of Section VI.)

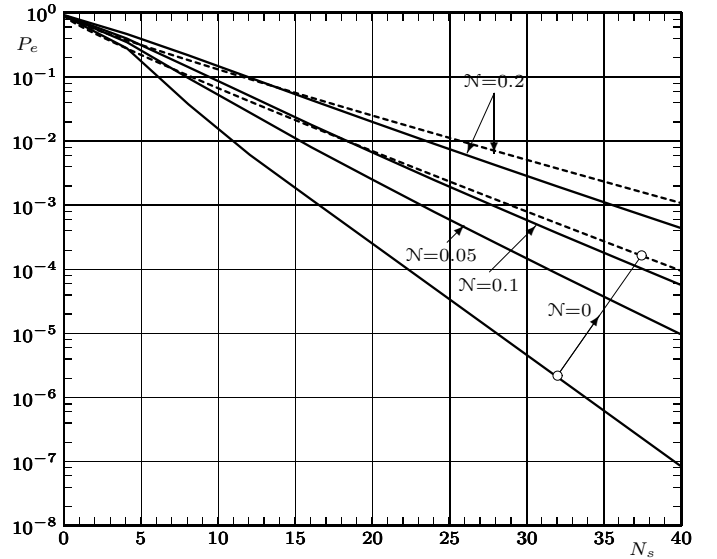

Fig. 2 — Error probability in 16–QAM versus $N_s$ for some values of $N$. Solid lines refer to quantum detection ans dashed lines to classical homodyne detection.

## VI. APPLICATION TO THE PSK MODULATION

The constellation of a coherent PSK modulation format is given by the Glauber states $|\gamma_i\rangle = |\gamma_0 W_m^i\rangle, i = 0,\ldots,m-1$, where, without loss of generality, we assume that $\gamma_0$ is real positive. This is a GUS constellation with initial state $|\gamma_0\rangle$ and

---

[2] $G$ depends on the four indexes $u, v, u', v'$, but it can be arranged as an ordinary matrix using the lexicographic order for the indexes.

[3] In Kato *et al.* [12] an analogous comparison is made with heterodyne detector.

generating operator

$$S = \sum_{n=0}^{\infty} W_m^n |n\rangle\langle n| \ . \tag{15}$$

With pure states the performance evaluation starts from the inner products $G_{0s} = \langle\gamma_0|\gamma_s\rangle$, which can be obtained from (8), namely

$$G_{0s} = \exp[-\gamma_0^2(1 - W_m^s)], \qquad s = 0, 1, \ldots, m-1 \ . \tag{16}$$

Then, the eigenvalues of the Gram matrix are obtained as the DFT of $G_{0s}$

$$D_k = \sum_{s=0}^{m-1} G_{0s} W_m^{-ks} \tag{17}$$

Finally, the minimum error probability is

$$P_e = 1 - \frac{1}{m^2}\left(\sum_{k=0}^{m-1}\sqrt{D_k}\right)^2 \ . \tag{18}$$

The above expressions are obtained without numerical approximations. $P_e$ is a function of the alphabet size $m$ and of the parameter $\gamma_0$, whose square $N_s = \gamma_0^2$ gives the *average number of photons per symbol*.

### A. Application of the SRM-GUS in the Presence of Noise

The $m$ Glauber density operators obtained with a PSK constellation verify the GUS with generating operator $S$ given by (15), which has infinite dimensions. In the SRM $S$ must be reduced to a finite dimension $n$ and then, in matrix form, it becomes

$$S = \mathrm{diag}\ [W_m^k, k = 0, 1, \ldots, n-1] \ .$$

Its eigendecomposition $S = Y\Lambda Y^*$ is trivial with $Y = I_n$ and $\Lambda = S$, and the matrices $L_k$ are given by

$$L_k = m\ \mathrm{diag}\ [\delta_{k, i \bmod m}, i = 0, 1, \ldots, n-1]$$

Now, for a given $n$, $N_s = \gamma_0^2$ and $\mathcal{N}$, the application of the SRM-GUS proceeds as follows: 1) evaluate the reference density operator $\rho_0 = \rho_0(\gamma_0)$ from (10); 2) find the factor $\gamma_0$ of $\rho_0$; 3) evaluate the blocks $D_k = \gamma_0^* L_k \gamma_0$ and find the square roots $D_k^{1/2}$ by eigendecomposition; 4) evaluate the blocks $B_{ii} = (G^{1/2})_{ii}$ from (6a); 5) evaluate $P_c$ from (6).

We give a detailed example of calculation in the case of small dimensions (for reason of space). We consider the 4-PSK with $N_s = 1$, $\mathcal{N} = 0.1$ and $N_e = 8$, which assures an accuracy of $\epsilon = 10^{-5}$. The reference density operator is the $8 \times 8$ matrix

$$\rho_0 = \begin{bmatrix} 0.366 & 0.333 & 0.214 & 0.112 & 0.051 & 0.021 & 0.008 & 0.003 \\ 0.333 & 0.336 & 0.237 & 0.136 & 0.067 & 0.029 & 0.012 & 0.004 \\ 0.214 & 0.237 & 0.183 & 0.114 & 0.060 & 0.028 & 0.012 & 0.005 \\ 0.112 & 0.136 & 0.114 & 0.076 & 0.044 & 0.022 & 0.010 & 0.004 \\ 0.051 & 0.067 & 0.060 & 0.044 & 0.027 & 0.014 & 0.007 & 0.003 \\ 0.021 & 0.029 & 0.028 & 0.022 & 0.014 & 0.008 & 0.004 & 0.002 \\ 0.008 & 0.012 & 0.012 & 0.010 & 0.007 & 0.004 & 0.002 & 0.001 \\ 0.003 & 0.004 & 0.005 & 0.004 & 0.003 & 0.002 & 0.001 & 0.001 \end{bmatrix}$$

Its practical rank is 5. From the eigendecomposition of $\rho_0$ we obtain the $8 \times 5$ factor

$$\gamma_0 = \begin{bmatrix} -0.289 & 0.087 & -0.019 & -0.003 & 0.000 \\ -0.289 & 0.000 & 0.019 & 0.006 & 0.001 \\ -0.204 & -0.062 & 0.013 & -0.002 & -0.002 \\ -0.118 & -0.071 & -0.008 & -0.005 & 0.000 \\ -0.059 & -0.053 & -0.019 & 0.000 & 0.001 \\ -0.026 & -0.032 & -0.019 & 0.004 & 0.000 \\ -0.011 & -0.016 & -0.013 & 0.006 & -0.001 \\ -0.004 & -0.007 & -0.008 & 0.005 & -0.002 \end{bmatrix}$$

The matrices $L_k$ are

$$L_0 = \begin{bmatrix} 4 & 0 & 0 & 0 & 0 & 0 & 0 & 0 \\ 0 & 0 & 0 & 0 & 0 & 0 & 0 & 0 \\ 0 & 0 & 0 & 0 & 0 & 0 & 0 & 0 \\ 0 & 0 & 0 & 0 & 0 & 0 & 0 & 0 \\ 0 & 0 & 0 & 0 & 4 & 0 & 0 & 0 \\ 0 & 0 & 0 & 0 & 0 & 0 & 0 & 0 \\ 0 & 0 & 0 & 0 & 0 & 0 & 0 & 0 \\ 0 & 0 & 0 & 0 & 0 & 0 & 0 & 0 \end{bmatrix} \qquad L_1 = \begin{bmatrix} 0 & 0 & 0 & 0 & 0 & 0 & 0 & 0 \\ 0 & 4 & 0 & 0 & 0 & 0 & 0 & 0 \\ 0 & 0 & 0 & 0 & 0 & 0 & 0 & 0 \\ 0 & 0 & 0 & 0 & 0 & 0 & 0 & 0 \\ 0 & 0 & 0 & 0 & 0 & 0 & 0 & 0 \\ 0 & 0 & 0 & 0 & 0 & 4 & 0 & 0 \\ 0 & 0 & 0 & 0 & 0 & 0 & 0 & 0 \\ 0 & 0 & 0 & 0 & 0 & 0 & 0 & 0 \end{bmatrix} \quad \ldots$$

Then the evaluation of $D_k$ and its square root, e.g. for $k = 0$, gives

$$D_0 = \begin{bmatrix} 0.348 & -0.088 & 0.026 & 0.004 & 0.000 \\ -0.088 & 0.042 & -0.002 & -0.001 & 0.000 \\ 0.026 & -0.002 & 0.003 & 0.000 & 0.000 \\ 0.004 & -0.001 & 0.000 & 0.000 & 0.000 \\ 0.000 & 0.000 & 0.000 & 0.000 & 0.000 \end{bmatrix}$$

$$D_0^{1/2} = \begin{bmatrix} 0.576 & -0.121 & 0.048 & 0.006 & 0.000 \\ -0.121 & 0.164 & 0.020 & -0.002 & -0.003 \\ 0.048 & 0.020 & 0.010 & 0.000 & -0.001 \\ 0.006 & -0.002 & 0.000 & 0.000 & 0.000 \\ 0.000 & -0.003 & -0.001 & 0.000 & 0.000 \end{bmatrix}$$

Finally, the probabilities are

$$p_c = \begin{bmatrix} 0.80703 & 0.08622 & 0.02034 & 0.08622 \\ 0.08622 & 0.80703 & 0.08622 & 0.02034 \\ 0.02034 & 0.08622 & 0.80703 & 0.08622 \\ 0.08622 & 0.02034 & 0.08622 & 0.80703 \end{bmatrix}$$

$$P_c = 0.80703 \qquad P_e = 0.19297 \ .$$

### B. Performance of 4-PSK and 8-PSK

The SRM-GUS approach has been applied to evaluate the error probability $P_e$ in 4-PSK and 8-PSK systems following the steps outlined above. The results are illustrated in Fig. 3 and in Fig. 4, where $P_e$ is plotted versus the average number of photons per symbol $N_s$ for some values of the thermal noise parameter $\mathcal{N}$. In particular, the curve for $\mathcal{N} = 0$, which refers to the absence of thermal noise, was checked with the results obtained with pure states (see (18)) and a perfect agreement has been found. To assure an overall accuracy of $\epsilon = 10^{-5}$, the reference density operator $\rho_0$ was approximated with a matrix of size $N_e = 145$ with a rank running from 1 to 48 in dependence of $\mathcal{N}$.

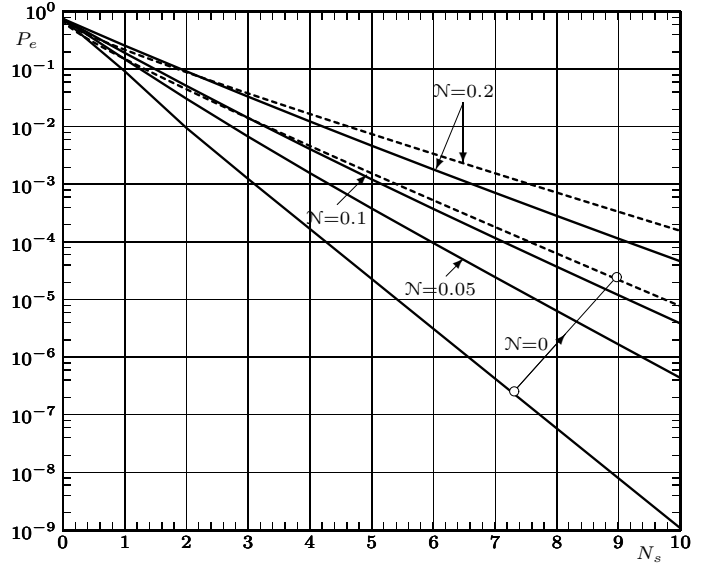

Fig. 3 — Error probability in 4-PSK versus $N_s$ for some values of $\mathcal{N}$. Solid lines refer to quantum detection and dashed lines to classical homodyne detection.

In Fig. 3 and Fig. 4 the PSK quantum detection is compared with homodyne counterpart. Remarks similar to that made for 16-QAM can be repeated.

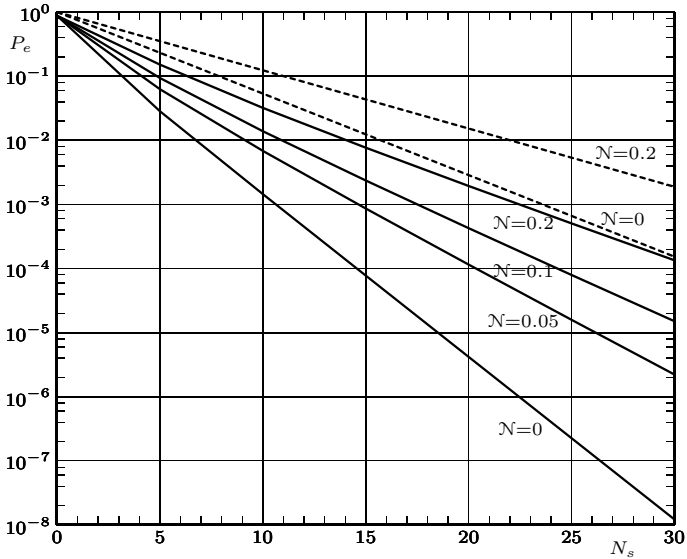

Fig. 4 — Error probability in 8–PSK versus $N_s$ for some values of $\mathcal{N}$. Solid lines refer to quantum detection and dashed lines to classical homodyne detection.

### C. SRM vs Optimal Measurement

For mixed states, the SRM approach is not optimal, at least in general, so that a comparison with the optimal performances is adequate. As mentioned above, the maximum probability of correct detection is the maximum of $\sum_{i=0}^{m-1} \mathrm{Tr}(q_i \rho_i \Pi_i)$ under the constraints that the $\Pi_i$ are PSD and resolve the identity $I_{\mathcal{H}}$ or, equivalently, the minimum of $\mathrm{Tr}(Y)$ under the constraint that the operators $Y - q_i \rho_i$ are PSD. This is a problem of convex semidefinite programming. The numerical evaluation of the optimum for the applications considered in the paper have been performed by Matlab with the LMI (Linear Matrix Inequality) Toolbox. A comparison is presented in Fig. 5 for the 2–PSK and 4–PSK systems and shows that the SRM exhibits an error probability about 30% greater than the optimum detector. Since similar results hold true also for 8–PSK and 16–QAM, we conclude that SRM is "pretty good" also in the presence of thermal noise.

**Remark.** For the 2–PSK (as for any other binary format) the exact evaluation can be carried out using Helstrom's theory (see [10]). This possibility was used to check the results obtained with the Matlab LMI toolbox.

## VII. CONCLUSIONS

We calculated the error probability of QAM and PSK quantum data communication system in the presence of thermal noise with quantum detection based on the SRM technique. The main novelty of the paper lies in the performance evaluation of quantum data transmission system affected by thermal noise, not necessarily in a small amount. The lack of results in the literature about this topic is surely due to the difficulties of numerical computation of optimal values. The extension of the SRM approach to mixed states by Eldar *et al.* [6] allowed us to develop such computations with a relatively limited amount of numerical complexity.

Comparisons made with the performance of classical homodyne detection give results similar to that for OOK and BPSK schemes and evidence the superiority of the quantum detection also in the presence of thermal noise. A comparison of the SRM with the optimal detection performance, evaluated by a convex semidefinite programming package, shows only a moderate impairment, so that the obtained results can be considered a very good approximation of the optimal performances.

The results of the application of the SRM to geometrically uniform symmetric states will enable one to consider other quantum modulation schemes, both in absence and in presence of thermal noise. In particular, pulse position modulation (PPM) has recently been considered [3] for possible applications to deep space quantum communications.

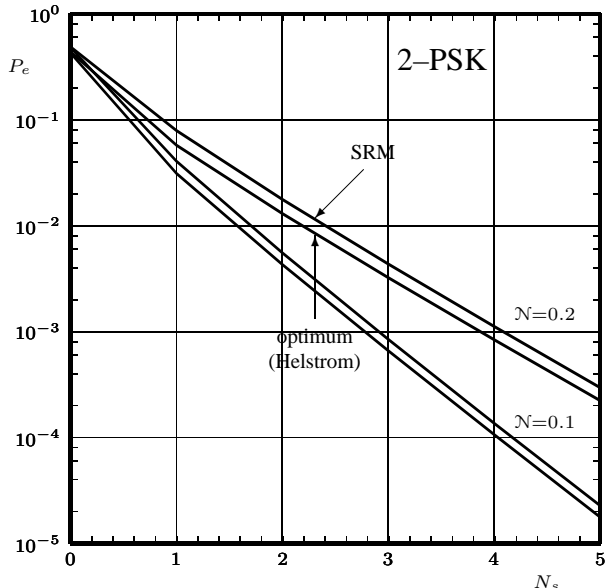

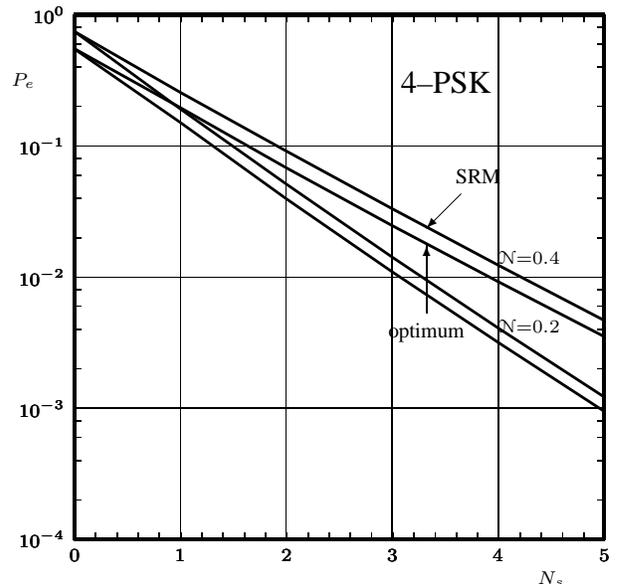

Fig. 5 — Error probability in 2–PSK and 4–PSK versus $N_s$ for some values of $\mathcal{N}$. The SRM detection is compared with the optimum detection.

However, further problems of computational complexity arise, owing to the fact that the natural model for the PPM scheme is the tensorial product of Hilbert spaces. Our research on the topic is in progress.

*Acknowledgment*

The authors would like to thank Roberto Corvaja for his helpful support and criticism.